\documentclass[aps,prl,reprint,twocolumn,superscriptaddress]{revtex4-2}

\usepackage{amsmath} 
\usepackage{amssymb} 
\usepackage{graphicx}
\usepackage{dcolumn}
\usepackage{bm}
\usepackage[hidelinks]{hyperref}
\usepackage{color} 
\usepackage{siunitx}
\usepackage{comment}
\usepackage[capitalise]{cleveref}
\crefname{section}{Sec.}{Secs.}%
\usepackage{soul,xcolor}

\usepackage[nearskip=0pt,farskip=0pt,captionskip=0pt,caption=false]{subfig}

\newcommand{\phantomsubfloat}[1]{{
    \captionsetup[subfigure]{labelformat=empty}
    ~\\[-1.6em]
    \subfloat[][]{#1}
}}

\crefrangelabelformat{figure}{#3#1#4--#5(\crefstripprefix{#1}{#2}#6}
\crefmultiformat{figure}{Figs.~#2#1\xdef\mycreffirstarg{#1}#3}{ and~#2({\crefstripprefix{\mycreffirstarg}{#1}}#3}{, #2({\crefstripprefix{\mycreffirstarg}{#1}}#3}{ and~#2({\crefstripprefix{\mycreffirstarg}{#1}}#3}

\newcommand{\fig}[1]{\cref{fig:#1}}

\newcommand{\eq}[1]{Eq.~\eqref{eq:#1}}
\newcommand{\SM}[1]{\cite{SM}~\cref{#1}}
\newcommand{\Sec}[1]{\cref{#1}}

\setstcolor{red}

\begin{document}


\title{Temporal light control in complex media through the singular value decomposition\\of the time-gated transmission matrix}

\author{Louisiane Devaud}
\email{louisiane.devaud@lkb.ens.fr}
\affiliation{Laboratoire Kastler Brossel, ENS-Université PSL, CNRS, Sorbonne Université, Collège de France, 24 rue Lhomond, 75005 Paris, France}

\author{Bernhard Rauer}
\affiliation{Laboratoire Kastler Brossel, ENS-Université PSL, CNRS, Sorbonne Université, Collège de France, 24 rue Lhomond, 75005 Paris, France}

\author{Matthias Kühmayer}
\affiliation{Institute for Theoretical Physics, Vienna University of Technology (TU Wien), A-1040 Vienna, Austria}

\author{Jakob\nolinebreak\ Melchard}
\affiliation{Institute for Theoretical Physics, Vienna University of Technology (TU Wien), A-1040 Vienna, Austria}

\author{Mickaël Mounaix}
\affiliation{Laboratoire Kastler Brossel, ENS-Université PSL, CNRS, Sorbonne Université, Collège de France, 24 rue Lhomond, 75005 Paris, France}
\affiliation{School of Information Technology and Electrical Engineering, The University of Queensland, Brisbane, QLD 4072, Australia}

\author{Stefan Rotter}
\affiliation{Institute for Theoretical Physics, Vienna University of Technology (TU Wien), A-1040 Vienna, Austria}

\author{Sylvain Gigan}
\affiliation{Laboratoire Kastler Brossel, ENS-Université PSL, CNRS, Sorbonne Université, Collège de France, 24 rue Lhomond, 75005 Paris, France}

\date{\today}

\begin{abstract}
The complex temporal behavior of an ultrashort pulse of light propagating through a multiply scattering medium can be characterized experimentally through a time-gated transmission matrix.
Using a spatial light modulator, we demonstrate here, that injecting singular vectors of this matrix allows us to optimally control energy deposition at any controllable delay time.
Our approach provides insights into fundamental aspects of multi-spectral light scattering and could find applications in imaging or coherent control.
\end{abstract}

\maketitle

A coherent light pulse passing through a multiple scattering medium gets distorted into speckled interference patterns---both in space and time~\cite{Goodman2007}.
Even though the resulting pattern is seemingly random, it results from linear and deterministic scattering events and can therefore be controlled~\cite{vellekoop2007focusing}.
In recent years, many techniques have been developed enabling the control of the scattered light by means of shaping the input field~\cite{mosk2012controlling,rotter2017light}. 

A concept that enables a particularly detailed control of light propagation through scattering media is the transmission matrix (TM)~\cite{popoff2010measuring}. 
For a given wavelength, the monochromatic TM relates the light field entering the medium to the output field, thus encoding the entire transmission behaviour of the scattering sample.
Once measured, an ``inversion'' of this matrix thus enables the focusing at arbitrary locations, the transmission of images~\cite{popoff2010image}, as well as the creation of states that are invariant to scattering~\cite{pai2021scattering}.
In addition, decomposing the matrix in its eigen- or singular modes gives access to orthogonal input channels sorted by their total transmission~\cite{kim2012maximal,yu2013measuring}.
This is not only interesting for the study of mesoscopic effects, accessible when measuring a large fraction of the modes, such as open and closed channels~\cite{dorokhov1996coexistence,pendry1990maximal,beenakker1997random}, but also for controlling the overall transmission and reflection of a sample~\cite{vellekoop2008universal,gerardin2014full,hsu2017correlation}.

In the case of pulsed illumination, the speckled interference pattern acquires an additional temporal component as different wavelengths of the pulse propagate differently in the scattering process~\cite{mccabe2011spatio}.
The TM concept still holds in this situation but needs to be extended to a multispectral TM, encoding the transmission of each spectral channel~\cite{andreoli2015deterministic}.
The number of distinct spectral channels needed to capture the full transmission behaviour depends on the scattering sample and on the bandwidth of the input pulse.
In order to address the light arriving at a certain time one can either Fourier-transform a multispectral TM \cite{xiong2019long,mounaix2019control} or directly perform a time-resolved measurement, resulting in a time-gated TM which is valid only for a certain arrival time in the scattered pulse.
This tool already proved useful for spatio-temporal focusing behind scattering media~\cite{mounaix2016deterministic}, for focusing on an object embedded inside a scattering sample~\cite{jeong2018focusing} and, more recently, to control the energy delivery after propagation in a multimode fiber~\cite{mounaix2019control,xiong2019long}.
The two latter studies, however, only indirectly measure and manipulate the pulse response by measuring its individual monochromatic components and by reconstructing it in post-processing.

In this Letter, we report on the temporal control of a femtosecond laser pulse that passes through a multiple scattering medium.
We characterize the scattering process through the direct measurement of the time-gated TM by interfering the scattered light with a controllable delayed probe pulse.
Performing the singular value decomposition (SVD) of this time-gated TM provides us with a set of input modes that enable temporal control, beyond simple focusing~\cite{mounaix2016deterministic}.
Through direct measurements of the time-resolved transmission, we are able to show that the singular vectors of the time-gated TM can be used to enhance or diminish the energy arriving at any selected time in the output pulse, enabling a smooth and accurate control of the temporal intensity distribution.
We complement our experimental observations with simulations of scattering waveguides that allow us to probe regimes of control not easily accessible experimentally.
In this way we uncover the existence of states that are perfectly non-transmitting at given delay times in the output pulse.
Working with the SVD has the advantage of accessing the state with \textit{optimal} transmission at any given time---a feature recently addressed in the context of monochromatic light~\cite{kim2012maximal}, broadband inputs in optical fibers~\cite{xiong2019long, mounaix2019control}, the deposition matrix~\cite{bender2022depth} or non-normal photonic media~\cite{makris2014anomalous}.
We illustrate this aspect by comparing the SVD states to other strategies to enhance the total transmission.
To illustrate the versatility of the SVD-based transmission control, we show how to use it for controlling the speckle grain size at a specific time~\cite{devaud2021speckle}.  

\begin{figure}[t]
    \includegraphics[width=0.95\columnwidth]{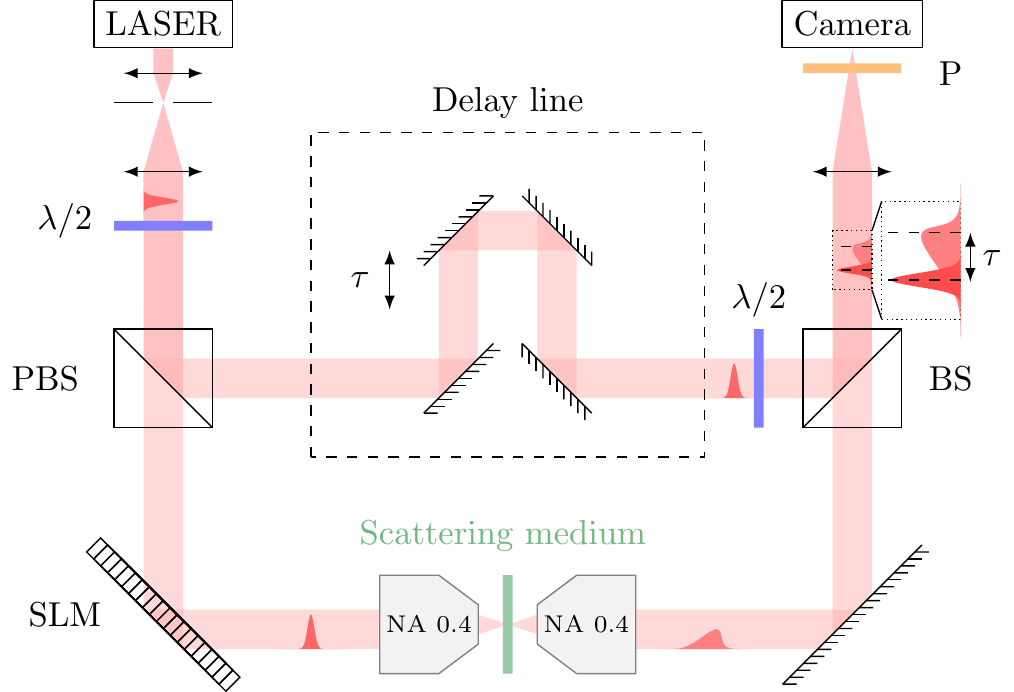}
    \caption{Scheme of the experimental setup.
    An ultrashort pulse of light, of central wavelength $\lambda_0 = \SI{808}{\nano \metre}$, delivered from a Ti:sapphire pulsed laser (MaiTai HP, Spectra Physics, $\simeq\SI{100}{\femto \second}$ pulse length) is divided upon two paths by a polarizing beam splitter (PBS).
    On one path the pulse wavefront is modulated by a reflective phase-only SLM (HSP512L-1064, Meadowlarks) and passes through a $\sim$\SI{10}{\micro \metre} scattering layer of TiO$_2$ particles (transmittance of $\sim 0.3$, suspended on a glass slide and static on the timescale of experiments) where it gets elongated.
    On the second path, the pulse is sent on a controlled delay line. 
    Both pulses get recombined on a beam splitter (BS) and are imaged on a CCD camera (Manta, G-046, Allied Vision).
    Two half-wave plates ($\lambda/2$) adjust the polarization of the power of both arms and a polarizer (P) before the camera selects the desired polarization.}
    \label{fig:experimental_setup}
\end{figure}

The experimental setup used is sketched in~\fig{experimental_setup}.
An ultrashort pulse of light, modulated by a reflective phase-only spatial light modulator (SLM) is sent through a slab of scattering material.
Multiple scattering processes elongate the pulse in time before it is imaged on a charged coupled device (CCD) camera.
Additionally, an unperturbed plane wave probe pulse, decoupled from the beam before the SLM, is recombined and interferes with the scattered light at the CCD. 
A delay line in the probe path allows us to tune the time delay $\tau$ between the scattered light and the probe pulse.
Scanning the probe pulse delay over the elongated pulse enables to retrieve the spatio-temporal field of the scattered light~\cite{yaqoob2008optical,mounaix2016deterministic} (for more details refer to~\SM{SM_field_measurement,SM_zero_delay}).

This time-gated configuration enables us to measure a time-gated TM at any desired delay time $\tau_0$ within the distribution of delays induced by the scattering.
To this end, a set of orthogonal spatial modes (Hadamard basis) is displayed on the SLM while for each mode the output field is measured on the CCD by phase-stepping the pattern on the SLM between 0 and $2\pi$~\cite{popoff2010measuring}.
As the probe pulse only interferes with the part of the scattered light that matches its delay, the measured TM only encodes information about this given delay.
The rest of the light acts as a background and does not influence the TM measurement.

\begin{figure}[t]
    \includegraphics[width=1\columnwidth]{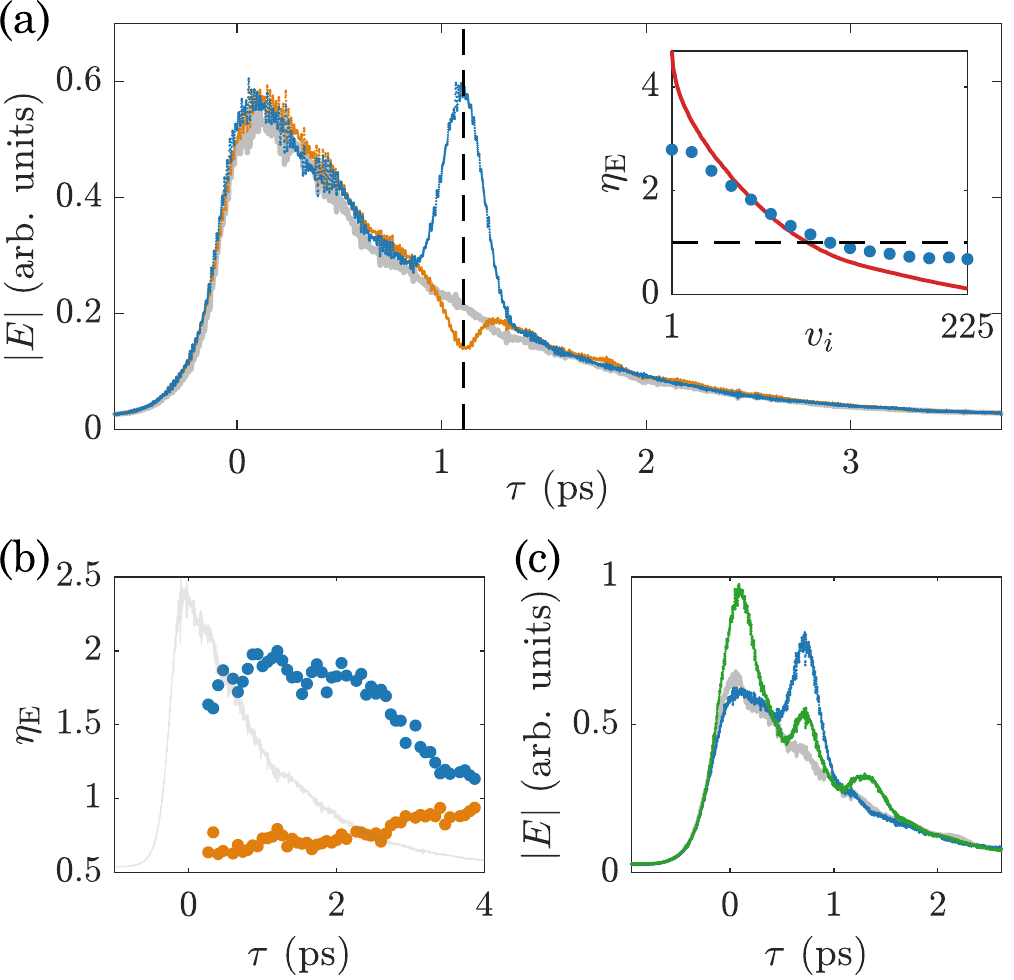}
    \phantomsubfloat{\label{fig:exp_results-a}}
    \phantomsubfloat{\label{fig:exp_results-b}}
    \phantomsubfloat{\label{fig:exp_results-c}}
    \caption{Temporal control of scattered light with the time-gated TM.
    (a) Temporal profile of the spatially-averaged output pulse amplitude $|E|$ for the first singular vector $v_1$ (blue) and the last singular vector $v_{225}$ (orange) of a time-gated TM measured at $\tau_0$ = \SI{1.1}{\pico \second}.
    The pulse shape obtained for a plane wave input is shown in gray.
    In the inset, the amplitude enhancement $\eta_\mathrm{E}$ (measured at $\tau_0$) relative to the plane wave input is shown for the whole range of singular vectors (blue dots).
    The expected field enhancement for phase and amplitude control $\tilde{s} = s / \sqrt{\langle s^2 \rangle}$ corresponds to the normalized singular values (see~\SM{SM_minimal_model}) and is indicated by the red solid line.
    (b) Enhancement obtained with the first singular vector for different delay times $\tau_0$ (blue dots).
    The plane wave output pulse is shown as a visual aid (arbitrary units, gray).
    Here, the individual time-gated TMs have been measured with 10 phase steps per mode (instead of 4, see~\SM{SM_field_measurement}) to reduce measurement noise in the pulse tail.
    (c) Simultaneous enhancement at three different delay times by projecting the sum of the corresponding TM's first singular vectors (green) compared to the plane wave input (gray) and to a single time control (blue).
    All TMs used here were measured for $N_{\mathrm{SLM}} \simeq 640$ (see~\SM{SM_Def_gamma}) and $N_{\mathrm{CCD}} = 225$.
    Data displayed in (a) are averaged over 4 disorder realizations whereas (b) and (c) correspond to a single realization.
    } 
    \label{fig:exp_results}
\end{figure}

Using phase conjugation, it has already been shown that the information contained in the TM can be used to concentrate light at single or multiple spatio-temporal positions~\cite{mounaix2016deterministic}. Here, instead, we address the question of how to achieve the globally optimal energy delivery in a selected output region, at a pre-determined time $\tau_0$.
The method of choice for this task is to perform an SVD of the time-gated TM, which yields the real singular values $s_i$ and the orthogonal input singular vectors $v_i$ associated with them (see~\SM{SM_Def}).
The number of non-zero singular values at a given time $\tau_0$ corresponds to the matrix rank and we sort them in a decreasing order.
Correspondingly, the first singular vector $v_1$ corresponds to the largest singular value and thus to the globally optimal transmission at the time delay $\tau_0$.
In the experiment, we display the phase of the singular vectors on the SLM and record the resulting output pulses.
The impact of a specific input wavefront on the temporal shape of the output is measured by tuning the delay of the probe pulse over the full extended duration of the scattered pulse while recording their interference.
In \cref{fig:exp_results-a} we present the temporal shape of the scattered field amplitude, spatially averaged over the full area over which the time-gated TM was measured for two of its singular vectors.
The first singular vector $v_1$, associated with the highest singular value, leads to a sharp increase of the field amplitude in the region of interest at $\tau_0$.
On the contrary, the last singular vector $v_{225}$ ($N_{\mathrm{CCD}} = 225$), associated with the lowest singular value, results in a decrease at that delay $\tau_0$.
Outside of the controlled time-bin, the temporal profile of the pulse is not affected.

These two singular vectors represent the extreme cases of field enhancement or reduction. 
Sending singular vectors associated to intermediate singular values enables us to tune the enhancement to field values in between, see the inset of \cref{fig:exp_results-a}.
Here, we define the amplitude increase ratio $\eta_{\mathrm{E}}$ as the relative enhancement, at the target time, of the field amplitude relative to the unshaped plane wave input pulse. 
The measured enhancement values follow qualitatively the TM singular values distribution.
The observed discrepancy in magnitude of the enhancement can be explained by the phase-only constraint of our wavefront modulation while the global shape is reproduced well by a random matrix model (see~\SM{SM_minimal_model}).
The singular vector spectrum of the time-gated TM hence enables a smooth control of the energy delivery at the  target time $\tau_0$.

Through measuring the time-gated TM for different delay times, temporal control can be gained over the whole duration of the elongated pulse, as shown in \cref{fig:exp_results-b} for the first singular vector.
Yet, its effectiveness depends on the value of $\tau_0$, with the increase ratio reaching a maximum at around $\tau_0 \approx \SI{1.1}{\pico\second}$ which corresponds about double the Thouless time of the medium~\cite{thouless1977maximum}.
The decay of the enhancement for late times can be attributed to a drop in the signal to background ratio of the TM measurement (in line with related observations in~\cite{aulbach2011control}).
Our work also shows that the delay at which the energy variations apply can slightly differ from $\tau_0$, especially for short and long delays (see~\SM{SM_increase_ratio}).

Up to now, we demonstrated control of the global pulse amplitude at a single delay time only.
However, to gain control over multiple times at once, linearity enables us to sum the singular vectors associated to time-gated TMs measured at different times, as presented in \cref{fig:exp_results-c}.
This is analogous to the superposition of multiple spatial, spatio-temporal or spatio-chromatic foci ~\cite{popoff2010measuring,mounaix2016deterministic,mounaix2016spatiotemporal,andreoli2015deterministic}.
As for the spatial superpositions, tuning the pulse amplitude at multiple times with a single SLM mask naturally leads to a drop in efficiency.

\begin{figure}[t]
    \includegraphics[width=0.95\columnwidth]{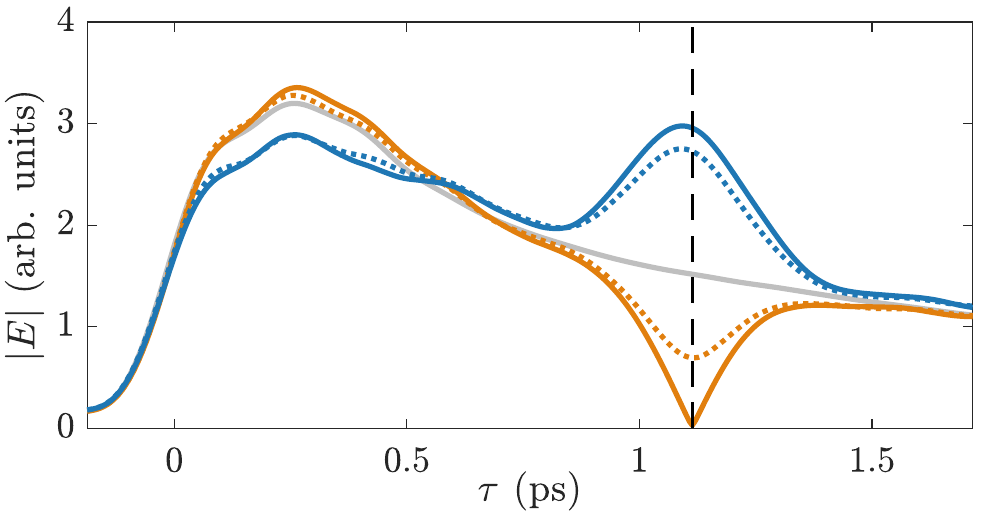}
    \caption{Simulation results in a waveguide geometry with a similar scattering strength and using 50 input and output modes.
    Solid lines indicate the output pulse shapes for amplitude and phase control of the singular vectors $v_1$ (blue) and $v_{50}$ (orange) corresponding to the largest and smallest singular value of the TM at the value of $\tau_0$ (indicated by the vertical dashed black line).
    The corresponding results for phase-only control are depicted by dashed lines.
    The gray line represents the time of flight reference when injecting a set of random inputs.
    The simulation results are averaged over 10 realizations of the disorder.
    }
    \label{fig:fig_theory}
\end{figure}

An important technical limitation in our setup is the restriction to phase-only input patterns.
Even though the singular vectors contain the amplitude information, we only project their phase on the SLM.
In order to better understand how this limitation affects the temporal shaping presented, we turn to simulations.
For a two-dimensional waveguide geometry filled with randomly placed obstacles we solve the scalar Helmholtz equation for 50 transverse input and output modes (for details see~\SM{SM_simu}).
Doing this for a range of wavelengths we can retrieve the temporal response of the waveguide through a Fourier transform~\cite{mounaix2019control,xiong2019long}.
This provides us with the time-resolved TM, allows us to calculate its SVD and to evaluate the output pulse shape for different input singular vectors.
\Cref{fig:fig_theory} shows the response for the extremal vectors $v_1$ and $v_{50}$, both for phase and amplitude and phase-only control.
For the latter, we find good qualitative agreement with the experimental results.
As expected, the additional amplitude control yields a larger modulation of the field at the target delay (indicated by the black dashed line).
In the case of the first singular vector, however, the difference between full control and phase-only control is not large.
For the last singular vector, on the other hand, the additional amplitude control allows us to create a zero crossing of the electric field (in line with results obtained in multimode fibers where a quasi-cancellation of the field was observed~\cite{mounaix2019control}).

Note that, in the waveguide, the fraction of modes that is controlled is limited to the lowest 50\% (see~\SM{SM_simu}), whereas in the experiment a much smaller fraction of all modes is controlled.
In addition, the transverse boundary conditions differ.
The good agreement between the simulation and the experimental results is therefore all the more striking and indicates that SVD-based temporal control is a widely applicable tool.

\begin{figure}[t]
    \includegraphics[width=1\columnwidth]{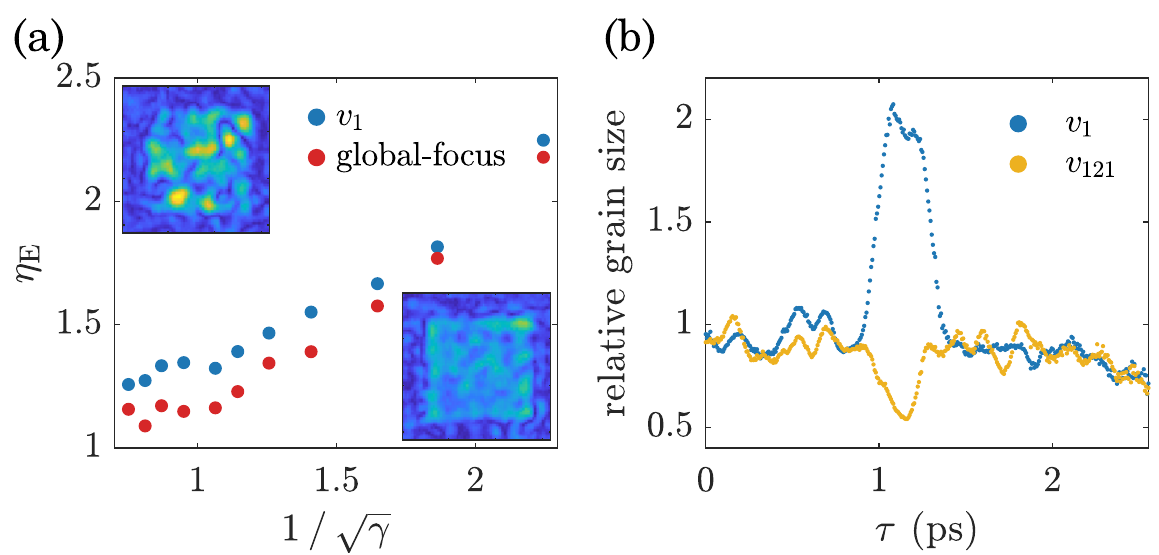}
    \phantomsubfloat{\label{fig:sum-a}}
    \phantomsubfloat{\label{fig:sum-b}}
    \caption{(a) Amplitude enhancement $\eta_{\mathrm{E}}$ over the degree of control scaled as $1/\sqrt{\gamma}$ for the first singular vector $v_1$ (blue) and the global-focusing vector (red).
    Here $N_{\mathrm{SLM}} \simeq 120$ was fixed and $N_{\mathrm{CCD}}$ was varied from $N_{\mathrm{CCD}}=25$ ($1/\sqrt{\gamma}=2.25$) to $N_{\mathrm{CCD}}=225$ ($1/\sqrt{\gamma}=0.75$).
    The insets show typical examples of the output field amplitude patterns for the first singular vector (top left) and the global-focusing vector (bottom right).
    Output fields are only affected in the region where the TM was measured, a central square, the edges are displayed for a visual reference.
    (No average over disorder realizations was performed.)
    (b) Temporal evolution of the speckle grain size for the first singular vector $v_1$ (blue) and an intermediate singular vector $v_{121}$ (yellow) in the same configuration used in \cref{fig:exp_results-a,fig:exp_results-b}.
    At each $\tau$ the speckle grain size is normalized to the one obtained for the plane wave input pulse.
    For all data presented in this figure the TM was measured at $\tau_0 =\SI{1.1}{\pico\second}$.
    (An average over 4 disorder realizations was performed.)
    }
    \label{fig:sum}
\end{figure}

In the following, we compare our SVD control to 
conceptually simpler strategies to enhance the energy at a certain time.
A spatio-temporal focus, for example, also leads to a temporal focus, albeit only at a single location~\cite{mounaix2016deterministic}.
For the light to be focused on several locations at once, many focusing wavefronts can be superposed at the cost of a lower enhancement for the individual foci.
Contrary to what one might expect, focusing simultaneously on each output pixel for which the TM was measured does not entirely diminish the overall enhancement of the delivered light (see~\SM{SM_minimal_model}).
This global-focus pattern can be obtained by simply summing the time-gated TM over all output elements.
A comparison of the field enhancement obtained from the global focus and the first singular vector for different degrees of control $\gamma = N_{\mathrm{CCD}}/N_{\mathrm{SLM}}$ is presented in \cref{fig:sum-a}.
We see that for a large number of output modes ($1/ \sqrt{\gamma} \leq 1$) the SVD has a distinct advantage while in the case of few output pixels ($1/\sqrt{\gamma} \gg 1$) the global-focus leads to similar temporal enhancements.
In the extreme case of a single output pixel the first singular vector and the global-focus pattern are naturally equivalent~\cite{vellekoop2008phase}.
The scaling of the enhancement with $1/\sqrt{\gamma}$ is predicted analytically, see~\SM{SM_minimal_analytics}.

The smaller enhancements of the global-focus input compared to the first singular vector is expected since the summation of all TM rows constrains the field on all output pixels to have the same phase and amplitude values.
This restriction prevents the optimally transmitting mode from being reached and leads to non-Rayleigh distributed output patterns.
The insets in \cref{fig:sum-a} illustrate this by comparing the global-focus output to the Rayleigh distributed output speckle obtained with the first singular vector (see also~\SM{SM_speckle_stat}).

To demonstrate the versatility of working with singular vectors of the time-gated TM, we show here how this approach can also be used to control speckle correlations at well defined moments in time. For this purpose we translate a technique originally introduced for oversampled monochromatic TMs~\cite{devaud2021speckle} to the temporal domain. 
Relying on an imbalance in the transmission of different spatial frequencies encoded in the TM, this technique enables the control of the speckle grain size through a selection of different singular vectors.
For the time-gated TM, this effect translates to the time domain, allowing for a temporal control of the speckle grains as shown in~\cref{fig:sum-b}.
The grain size variations over the pulse are presented for two singular vectors and in analogy to the monochromatic case they can be tuned continuously when utilizing the full spectrum of singular vectors. We expect these results to trigger more advanced temporal correlation engineering in complex media.

In conclusion, we present an experimental technique based on the SVD of the time-gated TM which enables the temporal control of the global light delivery on the region of interest through a complex scattering medium.
By selecting different singular vectors as inputs we can continuously tune the transmission of a laser pulse at desired arrival times.
Such pulse shaping capabilities might be useful for pump-probe experiments or for non-linear excitations through scattering environments.
Moreover, we observe in simulations that amplitude and phase control at the input should allow for a perfect cancellation of the output field at a desired point in time.
We also compare the SVD-based transmission enhancement with a global-focusing approach that also boosts transmission, but falls short in efficiency and yields different output speckle statistics.
The preservation of Rayleigh statistics with the SVD may be an asset for speckle based imaging techniques such as speckle-field digital holography microscopy~\cite{park2009speckle} or for blind structured illumination microscopy~\cite{mudry2012structured}.
An important point to stress is that, compared to previous studies that investigated temporal power enhancements in multi-mode fibers~\cite{xiong2019long,mounaix2019control}, we perform measurements directly in the time domain.
Instead of measuring the full monochromatic response of the system and calculating the anticipated temporal response by means of a Fourier transformation, we work with pulsed light and time-gated measurements.
This means we only need a minimal number of measurements to shape the temporal light distribution, allowing us to explore interactions with non-linear processes in the future.
Our results will further improve the understanding of the time-gated TM and the fundamental possibilities of spatio-temporal light shaping in complex scattering environments.

\vspace{5mm}

\begin{acknowledgments}
This project was funding by the European Research Council under the grant agreement No. 724473 (SMARTIES), the European Union's
Horizon 2020 research and innovation program
under the Marie Sk\l odowska-Curie grant agreement No.\ 888707 (DEEP3P) and the Austrian Science Fund (FWF) under Project No.\ P32300 (WAVELAND).
The computational results presented in this paper were achieved using the Vienna Scientific Cluster (VSC).
M.M. acknowledges funding from the Australian Research Council (DE210100934).
S.G. is a member of the Institut Universitaire de France.
\end{acknowledgments}

\medskip

\bibliography{TGTM}


\setcounter{secnumdepth}{3}
\clearpage
\onecolumngrid

\renewcommand{\thefigure}{S\arabic{figure}}
\renewcommand{\theequation}{S\arabic{equation}}
\setcounter{equation}{0}
\setcounter{figure}{0}

\begin{center}
  \LARGE
  \textbf{Supplemental Materials}
\end{center}

\section{Field measurement: Delay line scan vs. holography}
\label{SM_field_measurement}

We use two techniques to extract the field of the scattered light.
The first relies on standard digital phase-stepping holography.
For this the delay line is fixed to the desired delay time of the probe pulse.
We then globally modulate the phase with the SLM (phase-stepping) and record the interference pattern between the scattered light and the plane wave probe pulse on the CCD.
From the different images recorded we retrieve the field of the scattered light.
We employ this technique to measure the transmission matrix as well as to retrieve the field for given fixed delays.
This technique is fast and especially useful for repeated measurement to allow for averaging.

The second technique relies on interferometric cross-correlation \cite{monmayrant2010newcomer}, which retrieves the entire temporal evolution of the scattered light within the pulse.
It can be seen as a continuous version of the phase-stepping holography described above.
Here the SLM displays a fixed pattern and the phase modulation is realised by tuning the delay line.
The delayed probe pulse is scanned over the broad scattered pulse while the CCD continuously records the resulting intereferogram.
Looking at a single pixel, the main frequency of the recorded signal is the carrier frequency of the laser.
Fourier filtering to extract just the amplitude of this oscillation returns the local pulse shape of the scattered light impinging on this pixel.
All temporal data presented in this article is measured in this way.
Compared to the first technique, this method is however quite slow such that in situations where repeated measurements are necessary, it becomes impractical.

\section{Definition of the temporal origin}
\label{SM_zero_delay}

\begin{figure}[h!]
    \phantomsubfloat{\label{fig:tau_zero-a}}
    \phantomsubfloat{\label{fig:tau_zero-b}}
    \includegraphics[width=0.7\columnwidth]{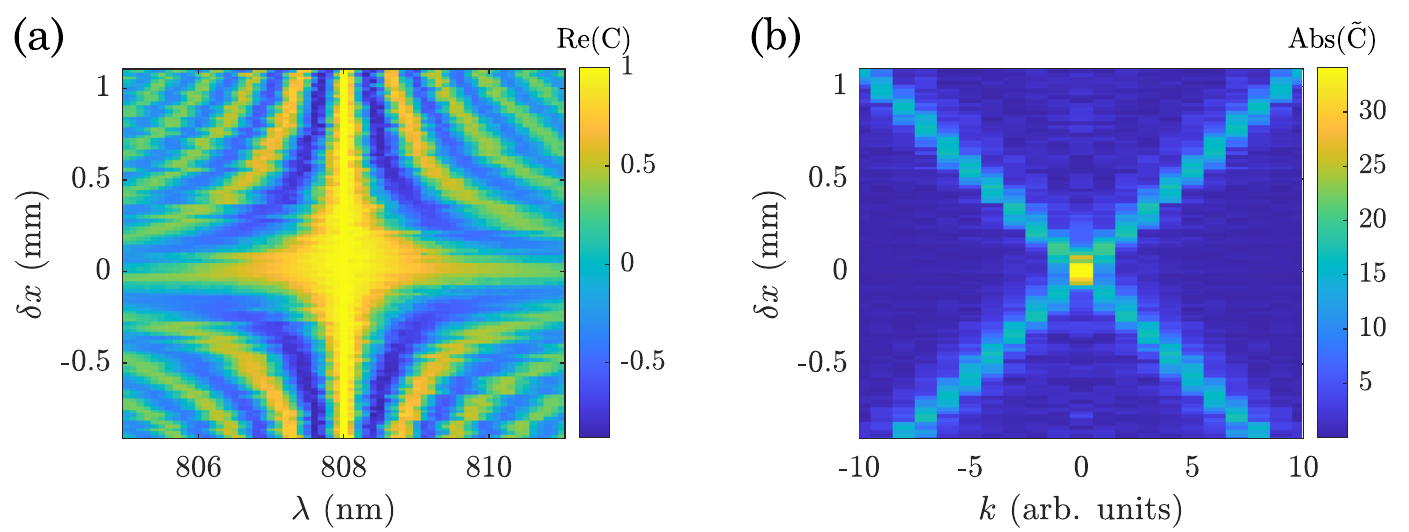}
    \caption{Measurement of the point of equal path length.
    (a) Real part of the correlation $C$ between the field speckle pattern obtained for a random input at wavelength $\lambda$ with the respective reference pattern at $\lambda_0 = \SI{808}{\nano\meter}$ for different lengths of the delay line $\delta x$.
    The fading of the pattern wavelength far from $\lambda_0$ results from the limited spectral decorrelation length of the medium.
    (b) Fourier transform $\tilde{C}$ along $\lambda$ of the data presented in (a).
    The crossing point of the oscillation peaks is used to determine $\delta x = 0$.
    }
    \label{fig:tau_zero}
\end{figure}

For the measured temporal data presented in \cref{fig:exp_results} as well as \cref{fig:sum-b} of the main text, the time $\tau = 0$ is defined to correspond to the delay line position $\delta x$ at which the two arms of the interferometer have the same optical path length in presence of the scattering medium.
To determine this point we use quasi-monochromatic light (pulse width of $\Delta\lambda < \SI{0.1}{\nano\meter}$) whose wavelength we tune continuously over a scale of $\pm \SI{3}{\nano\meter}$ around $\lambda_0 = \SI{808}{\nano \metre}$.
In the case of a non-zero path length difference $\lvert\delta x\rvert > 0$, different wavelengths will pick up different relative phases over $\delta x$.
When tuning the wavelength, this leads to an oscillation of the global phase of the field measured with the phase-stepping holography technique described above.
The frequency of this oscillation depends linearly on the path length difference and goes to zero when the two paths are of equal length.
We probe these oscillations by calculating the correlations between the measured $\lambda$-dependent field and the field measured at $\lambda_0$.
Doing this for different delay line positions $\delta x$ we obtain the pattern shown in \cref{fig:tau_zero-a}.
A clear symmetry point is observed when the path length  difference, and with it the oscillation frequency, goes to zero.
Fourier transforming the oscillations along $\lambda$ allows to determine $\delta x = 0$, defining the temporal origin (\cref{fig:tau_zero-b}).

For the simulations presented in \cref{fig:fig_theory} of the main text, the point $\tau = 0$ is defined as the time at which an input pulse would reach the output surface if the scattering region would be homogeneously filled with an averaged refractive index medium.
The propagation delay can be computed from the mean group velocity of the excited modes, see~\Sec{SM_simu}.

Note that the experimental method for determining $\delta x = 0$ is just an operational definition.
Also for monochromatic light a distribution of different path length through the medium exists which leads to the pulse shape affecting the observed oscillations.
However, we found that in practice this definition corresponds well with the temporal origin of the simulations.
In the end, the location of the temporal origin is arbitrary and does not play a role in the interpretation of the data.

\section{Definitions of characteristic quantities}
\label{SM_Def}

\subsection{Definition of $\gamma$}
\label{SM_Def_gamma}

\begin{figure}[h!]
    \phantomsubfloat{\label{fig:SM_nbr_SLM_modes-a}}
    \phantomsubfloat{\label{fig:SM_nbr_SLM_modes-b}}
    \includegraphics[width=0.6\columnwidth]{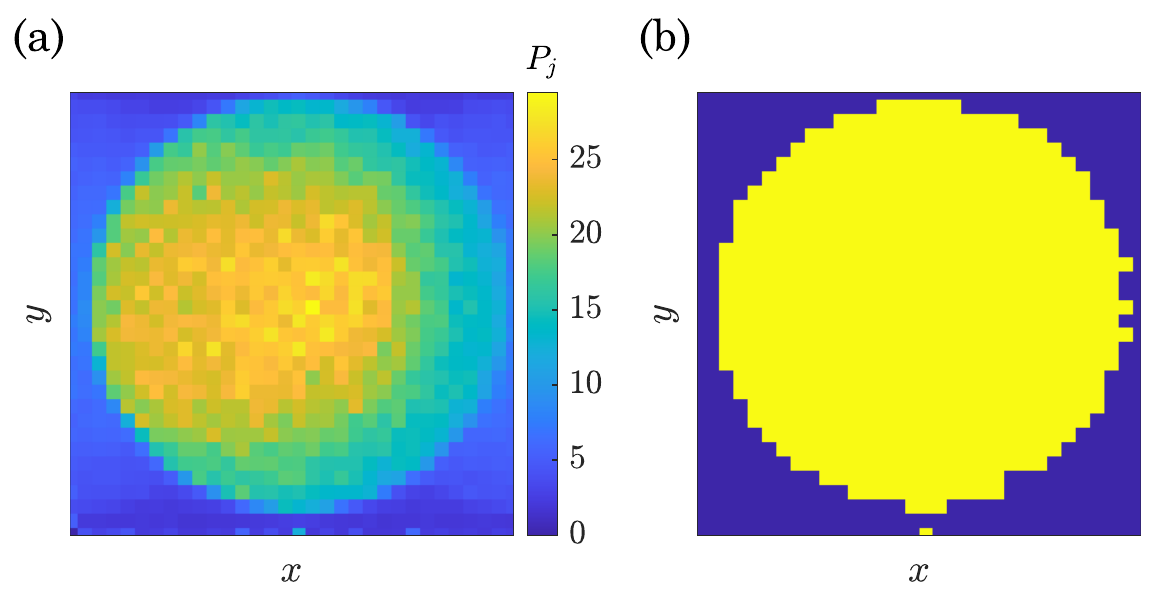}
    \caption{Effective number of SLM modes.
    (a) Image of the SLM modes contribution to the TM obtained by reshaping the vector $P$ obtained from \eq{SLM_modes}.
    (b) A threshold is applied and the number of effective modes computed by summing the modes above the threshold (fixed at 12 here).
    For the data in \fig{exp_results}(a) of the main text, this results in $N_{\mathrm{SLM}} \approx 640$ modes effectively launched to measure the TM.
    }
    \label{fig:SM_nbr_SLM_modes}
\end{figure}

\noindent
The amount of control in the experiment is defined by the parameter $\gamma$.
It is given by
\begin{equation}
    \gamma \equiv \frac{N_{\mathrm{out}}}{N_{\mathrm{in}}} \approx \frac{N_{\mathrm{CCD}}}{N_{\mathrm{SLM}}},
\end{equation}
where $N_{\mathrm{SLM}}$ is the number of modes controlled on the SLM and $N_{\mathrm{CCD}}$ the number of pixels in the region of interest (ROI) of the camera.
When measuring a TM the camera pixels are binned such that one pixel corresponds to one speckle grain.
The typical numbers of targeted modes used in this article are $N_{\mathrm{SLM}}^{\mathrm{target}}$~=~$32^2$ ($N_{\mathrm{SLM}}^{\mathrm{target}}$~=~$16^2$ and $N_{\mathrm{SLM}}^{\mathrm{target}}$~=~$62^2$ are used as well).
However, due to geometric experimental  limitations, i.e., the back focal plane of the illumination microscope objective cutting some SLM modes, the effective number of controlled modes needs to be estimated.
This is done using the information contained in the TM.
Taking the square root of the sum over the CCD dimension of the Hadamard product of the TM with its conjugate gives a vector that contains the information on the contribution of each SLM mode.
Let us denote $P$ the vector containing the information, one has
\begin{equation}
P_{j} = \sqrt{ \sum_i (T \cdot T^{*})_{i,j} } = \sqrt{ \sum_i T_{i,j} \times T^{*}_{i,j} },
\label{eq:SLM_modes}
\end{equation}
with $(A \cdot B)$ being the Hadamard product of two matrices $A$ and $B$ with equal dimensions.
The vector $P$ can be reshaped to visualise the SLM modes as presented in~\cref{fig:SM_nbr_SLM_modes-a}.
Applying a threshold and summing the number of modes above this threshold enables to obtain an estimate of the effective value of $N_{\mathrm{SLM}}$ (see~\cref{fig:SM_nbr_SLM_modes-b}).

The ROI on the camera is usually chosen large enough that we obtain smooth averaged pulse shapes while still being small enough that the probe pulse is sufficiently homogeneous over the whole area.
In most experiments we work with $\gamma$~=~0.2-0.3.
The speckle grain size is extracted by taking the half width at half maximum of the cross-correlation of the field.
To be insensitive to the sampling of the speckles this width is extracted by a Gaussian fit.
The speckle grain size can also be extrapolated from the TM measurement~\cite{davy2012focusing}.
It is given by the inverse of the participation number normalized by the rank of the TM.

\subsection{Definition of the normalized singular values}

To observe the Marchenko-Pastur law in the distribution of singular values of a TM they need to be normalized.
For a matrix of size $m \times n$ and singular values $s_i$ this normalization is usually defined as
\begin{equation}
    \tilde{s}_i^{\mathrm{MP}} = \frac{s_i}{\sqrt{\frac{1}{\mathrm{min}(n,m)}\sum_{j = 1}^{\mathrm{min}(n,m)}} s_i^2},
\end{equation}
regardless of the respective values of $m$ and $n$.
In physics normalizing the singular values of a TM such that they follow the Marchenko-Pastur law is often interesting~\cite{popoff2010measuring}.
However for a TM $m$ and $n$ have a physical meaning: the number of controlled modes ($N_{\mathrm{SLM}}$) and the degrees of freedom ($N_{\mathrm{CCD}}$).
They no longer are interchangeable.
To compute the mean over the singular values, all singular values equal to 0 should be included as they bring information on the transmission.
Here we hence will use an alternative version of the normalized singular values to be able to link them to the enhancement.
We define
\begin{equation}
    \tilde{s}_i = \frac{s_i}{\sqrt{\frac{1}{N_{\mathrm{SLM}}}\sum_{j = 1}^{N_{\mathrm{SLM}}}} s_i^2}.
\end{equation}
It is noteworthy that for $N_{\mathrm{SLM}} > N_{\mathrm{CCD}}$ (our experimental case), one has $\tilde{s}_i = \tilde{s}_i^{\mathrm{MP}}/{\sqrt{\gamma}}$.

\section{Temporal shift of the pulse modulation}
\label{SM_increase_ratio}

\begin{figure}[h]
    \phantomsubfloat{\label{fig:enhancement-a}}
    \phantomsubfloat{\label{fig:enhancement-b}}
    \includegraphics[width=0.6\columnwidth]{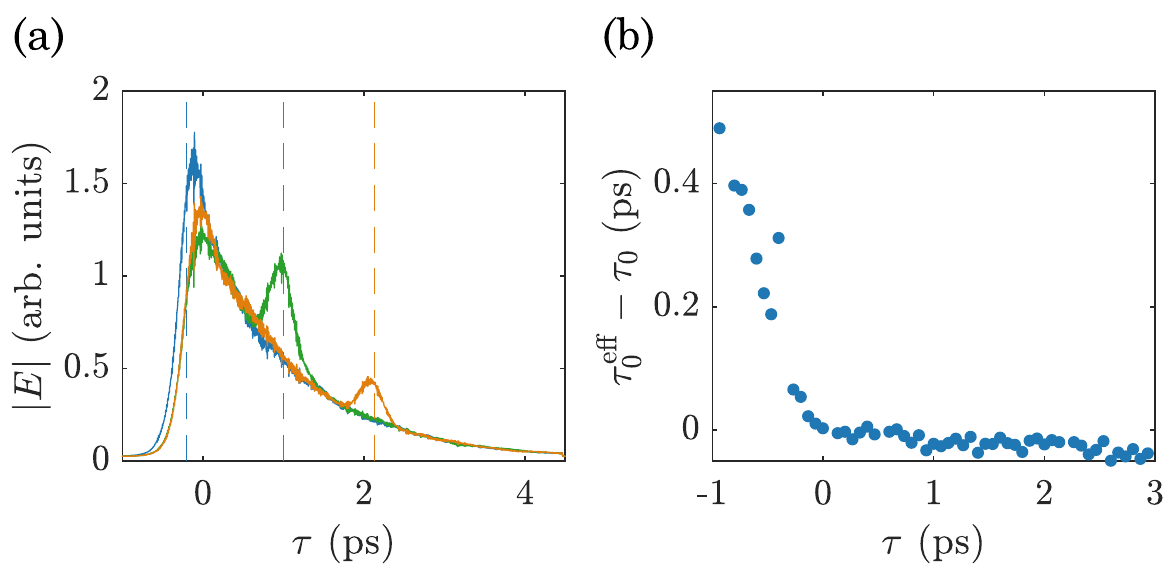}
    \caption{Evolution of the peak position over the pulse.
    (a) Pulse shape of the first SVD states $v_1$ for different values of $\tau_0$.
    It is possible to control energy delivery at all times within the distribution of delays induced by the scattering, but the peak positions do not exactly coincide with the chosen values of $\tau_0$ (indicated by the vertical dashed lines).
    At early times (in the pulse peak rise, blue curve) the increase is shifted to the later times, whereas at later times (pulse tail, orange curve) the peak is shifted to the earlier times.
    In both cases this corresponds to a shift in direction of the higher field.
    For intermediate delays ($\tau_0 \sim \SI{1}{\pico \second}$, green curve) no clear shift is visible.
    (b) Plot of the peak shift, given by $\tau_0^{\mathrm{eff}} - \tau_0$ where $\tau_0^{\mathrm{eff}}$ is the delay at which the peak is measured, relative to the $\tau_0$ value along the pulse.
    }
    \label{fig:enhancement}
\end{figure}

In the main text we show that the SVD of the time-gated TM allows us to modulate the transmitted amplitude at any given point in the pulse (see~\cref{fig:exp_results-c}).
Here, we want to point out that for early and late times the targeted enhancement (or reduction) does not appear exactly at the desired time, as shown in \cref{fig:enhancement-a,fig:enhancement-b}.
A way to understand this behaviour is to recall how the TM measurement is performed.
The probe pulse interferes with the elongated one at the chosen time $\tau_0$.
The probe pulse width is the one of the non-elongated pulse (full width at hall maximum \SI{100}{\femto \second}) so that the measurement is not temporally sharp but is multiplied by the Gaussian envelope of the probe.
There is then a temporal ``freedom'' for the peak position around the position $\tau_0$.
This explains the peak shifts to times at which it is easier to increase energy (i.e., where the field amplitude is higher).
This temporal position mismatch is taken into account in the enhancement value extraction: the enhancement is measured at the peak position instead of $\tau_0$.

\section{Simulation details}
\label{SM_simu}

In the numerical simulations, we solve the scalar Helmholtz equation $[\Delta + n^2(\vec{r}\,) k^2] \psi (\vec{r}\,) = 0$ in two dimensions on a regular Cartesian grid via the modular recursive Green's function method \cite{rotter2000greens,libisch2012graphene}.
Here, $\Delta = \partial_x^2 + \partial_y^2$ is the Laplacian in two dimensions and $n(\vec{r}\,)$ is the spatially-dependent refractive index distribution with $\vec{r} = (x,y)$ being the position vector.
Furthermore, $k = 2\pi/\lambda$ is the free space wave vector and $\psi(\vec{r}\,)$ is the unknown solution.

\begin{figure}
    \includegraphics[width=0.45\columnwidth]{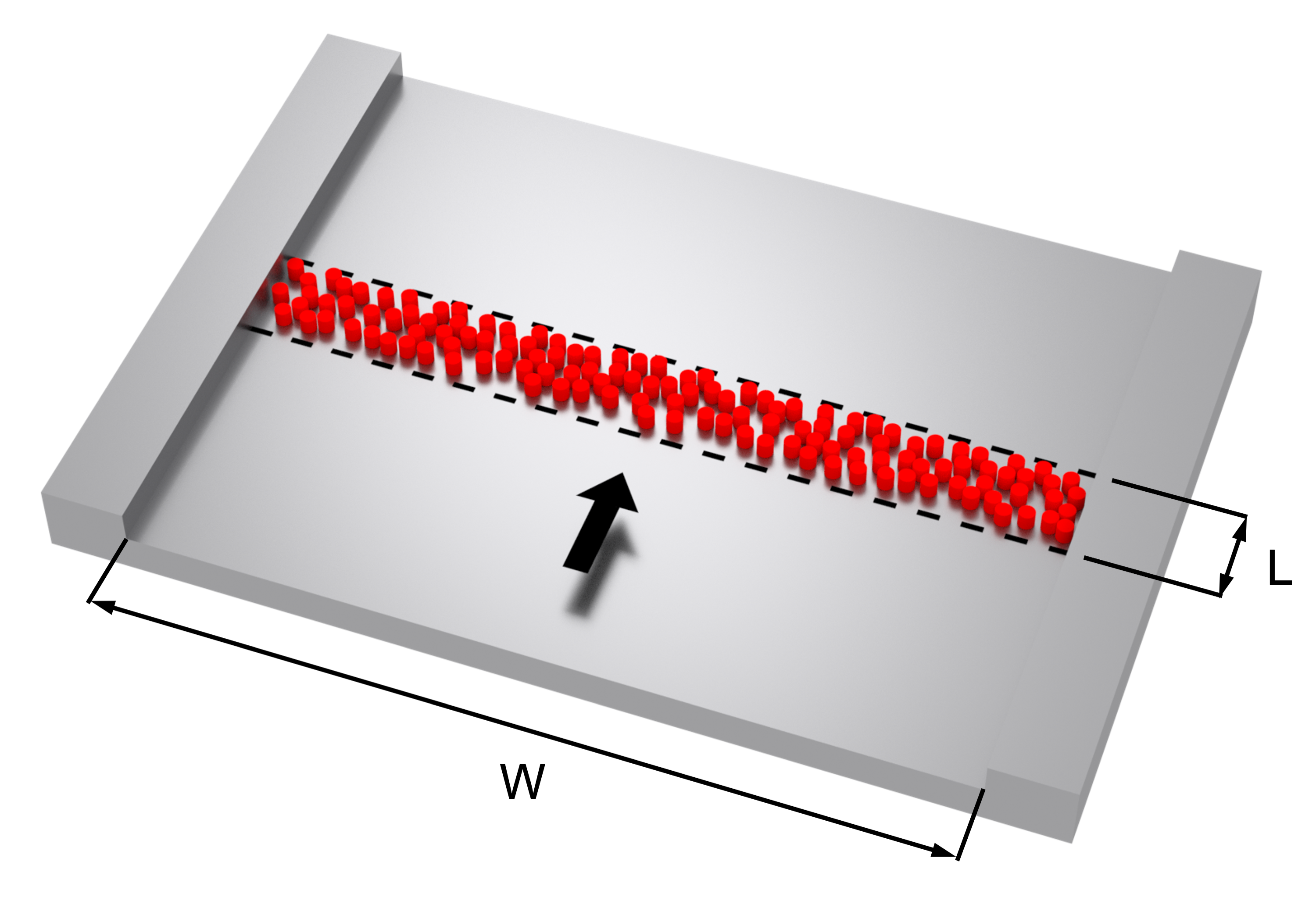}
    \caption{Sketch of the waveguide setup used in the numerical simulations.
    The scattering region with width $W = 1$ and length $L = W/10$ (whose boundary is marked by dashed black lines) contains circular obstacles (shown in red) with a radius of $R = W/100$ and refractive index of $n_\mathrm{scat} = 3.5$ that fill 40\% of its area.
    The black arrow marks the input side of the system.}
    \label{fig:SM_waveguide}
\end{figure}

The scattering system consists of a waveguide with a rectangular, slab-like scattering region (see~\cref{fig:SM_waveguide}) of width $W = 1$ and length $L = W/10$ in which circular obstacles of radius $R = W/100$ and refractive index $n_\mathrm{scat} = 3.5$ are placed.
To match the scattering strength of the experimentally used scattering samples, we use a filling fraction of $f_\mathrm{scat} = 0.4$ resulting in an average transmittance of $\sim 0.28$ (averaged over 10 disorder realizations) and a transport mean free path of $\ell_t \sim 0.31 L$ at the central frequency $\omega_0 = 75.55 c \pi/W$.
We then solve the monochromatic scattering problem for 301 frequencies in the interval $\omega \in [50.55, 100.55] c \pi / W$ to obtain the frequency-resolved transmission matrices $T(\omega)$ in the waveguide mode basis.

To arrive at the time-resolved transmission matrices $T(t)$, we Fourier-transform the frequency-resolved transmission matrices and only consider the lowest 50 waveguide modes at the input and output in order to avoid contributions from modes that are evanescent at certain frequencies.
Furthermore, we add a spectral function $f(\omega)$ to the Fourier transformation that defines the pulse shape which we choose to be a Gaussian.
More precisely, we use $f(\omega) = e^{-(\omega-\omega_0)^2/2\sigma_\omega^2}$ with $\sigma_\omega = \sqrt{2} \times 8/\langle \tau \rangle$.
Here, $\langle \tau \rangle = \pi A/C $ is the average time-delay in the scattering medium in two dimensions with $A = L W (1-f_\mathrm{scat}) + L W  f_\mathrm{scat} n_\mathrm{scat}^2$ corresponding to the area of the scattering region (the area of the dielectric scatterers has to be multiplied by their refractive index squared to account for the increased density of states) and $C = 2W$ being the external boundaries through which the waves can enter and exit the system \cite{pierrat2014pathlengthinvar,davy2021pathlengthinvar}.

In analogy to the experiment, we define the temporal origin $\tau = 0$ as the effective time-delay in a homogeneous medium with the same effective refractive index as the scattering medium (see~\Sec{SM_zero_delay}). The latter is given by $\tau_\mathrm{eff} = L / \langle v_g \rangle$, where $\langle v_g \rangle = (c/n_\mathrm{eff}) \langle k_x \rangle / \langle k \rangle$ is the mean group velocity with $c$ being the vacuum speed of light and $n_\mathrm{eff} = (1-f_\mathrm{scat}) + f_\mathrm{scat} n_\mathrm{scat}$ being the homogeneous effective refractive index. Moreover, $\langle k_x \rangle = \langle [ \langle k \rangle^2 - k_{y,n}^2 ]^{1/2} \rangle$ is the mode-averaged longitudinal propagation constant at the mean total wave vector $\langle k \rangle$ and $k_{y,n} = n \pi/W$ are the transverse wave vectors of the waveguide modes. In~\cref{fig:fig_theory} of the main text, we use a target time of $\tau_0 = 1.506 \langle \tau \rangle$, where the factor 1.506 has been chosen to match the position of the focusing peak with that in the experimental output pulse (at $\tau_0 = 1.1$~ps). All presented results are averaged over 10 disorder realizations with the same parameters.

\section{Minimal model to complement experimental observations}
\label{SM_minimal_model}

\subsection{Simulations}
\label{SM_minimal_model_simu}

\begin{figure}[h!]
    \phantomsubfloat{\label{fig:SM_minimal_model-a}}
    \phantomsubfloat{\label{fig:SM_minimal_model-b}}
    \includegraphics[width=0.7\columnwidth]{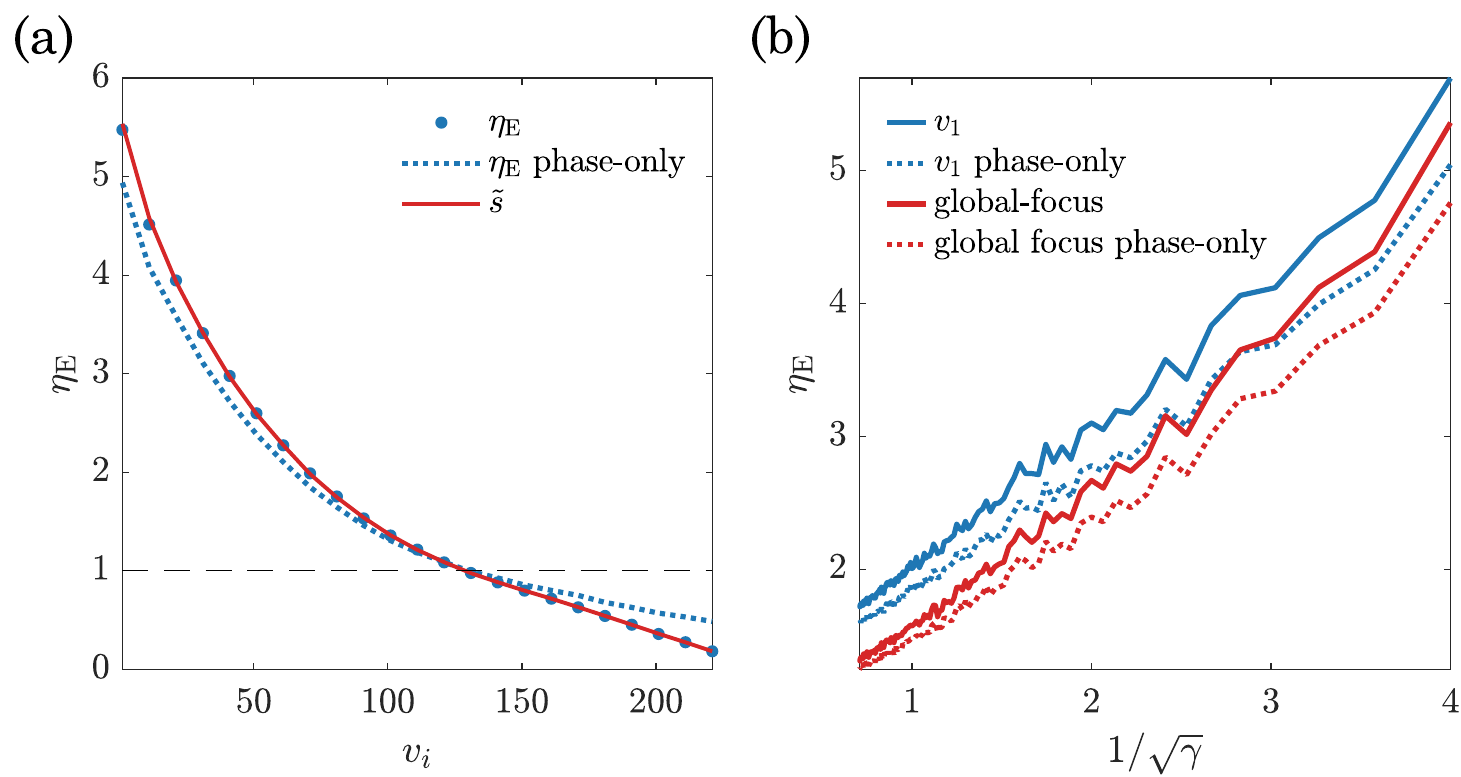}
    \caption{Minimal model results.
    (a) Field enhancement values obtained for different singular vectors in case of full control (blue dots) or phase-only control (blue dotted line).
    These values are plotted together with the normalized singular values $\tilde{s}$ (red line).
    The simulated TM is of size $1024 \times 225$ as for the measurements presented in~\fig{exp_results}(a) of the main text.
    The measured speckle grain size (1.7 pixels) is also accounted for in the simulation (see Supplemental Materials in~\cite{devaud2021speckle})).
    The data are averaged over 10 realizations of the disorder.
    (b) Comparison of the field enhancement values of the first singular vector $v_1$ (blue) and the global-focusing vector (red) for different degrees of control $\gamma$.
    The full control case is plotted with solid lines while the phase-only case is plotted with dotted lines.
    As for the experiment presented in \fig{sum}(a) of the main text, we vary $\gamma$ by varying the number of considered pixels in the ROI while keeping the number of SLM modes fixed at 256.
    Also here the experimental speckle grain size (1 pixel) is accounted for in the simulation.
    The data are averaged over 10 realizations of the disorder.
    }
    \label{fig:SM_minimal_model}
\end{figure}

We present the simulation results obtained from a minimal model in which the time-gated TM is regarded as a mere, numerically generated, random matrix (with complex Gaussian independent and identically distributed elements).
For this random TM we compute the output fields obtained for different input vectors, the singular modes or the global-focus input, and compare them with experimental observations.
In \cref{fig:SM_minimal_model-a}, the normalized singular values are compared to the field enhancement obtained in case of phase and amplitude control or phase-only control.
The values match well for full control.
In case of phase-only control, however, control over the output field is weaker both for increase or decrease, resulting in $\eta_E$ moving closer to 1.
\Cref{fig:SM_minimal_model-b} shows the evolution of the enhancement for the first singular vector and the global-focus vector with the degree of control $\gamma$.
As expected from the first singular vector being optimal, its enhancement is always higher than the one obtained for the global-focus. 
For relatively square TM (small $1/\gamma$), the effect is clearly visible.
The more non-square the TM gets the less difference there is in the observed enhancement.
Indeed, in the extreme case of only one non-zero singular value, its associated vector is the same as the global-focus one as there exists only a single output mode.
As expected, in the case of phase-only control the observed enhancements decrease.

\subsection{Analytical prediction}
\label{SM_minimal_analytics}

Experimentally, as shown in \fig{sum}(a) of the main text, and in the simulations presented in \fig{SM_minimal_model} one can observe that the first singular vector gives better enhancement results than the global-focusing input.
Here we want to analyse this difference analytically.
To do so let us consider a TM $T$ of size $n \times m$ (matrix dimensions given in subscript brackets) and its SVD: $T_{(n,m)} = U_{(n,n)} \times S_{(m,m)} \times V_{(m,m)} ^{\dagger}$.
The global-focusing vector $G_{(m,1)}$ is defined such that $G_{(m,1)} = T_{(m,n)}^{\dagger} I_{(n,1)}$ where the coefficients of I are all unity.
Now let us decompose $G$ in the basis of the singular vectors of $T$,
\begin{equation}
G_{(m,1)} = T_{(m,n)}^{\dagger} I_{(n,1)} = V_{(m,m)} S^{\dagger}_{(m,n)} U^{\dagger}_{(n,n)} I_{(n,1)} = \sum_i^{m} s_i \sum_j^{n} u^*_{j,i} V_i,
\end{equation}
where $s_i$ are the singular values and $u_{j,i}$ the elements of $U$.
The vector actually displayed on the SLM is normalized such that we have a field at the output $E_{\mathrm{G}}$:
\begin{equation}
    \tilde{G} = \frac{G}{||G||_2} = \frac{\sum_i s_i \sum_j u^*_{j,i} V_i}{\sqrt{\sum_i|s_i\sum_j u^*_{j,i}}|^2} \to E_{\mathrm{G}} = \frac{\sum_i s_i^2 \sum_j u^*_{j,i} U_i}{\sqrt{\sum_i|s_i\sum_j u^*_{j,i}}|^2}
\end{equation}
Similarly, when sending in a normalized random input $\tilde{R}$ one gets the field $E_{\mathrm{R}}$:
\begin{equation}
     \tilde{R} = \frac{\sum_i \beta_i V_i}{\sqrt{\sum_i|\beta_i|^2}} \to E_{\mathrm{R}} = \frac{\sum_i s_i \beta_i U_i}{\sqrt{\sum_i|\beta_i|^2}},
\end{equation}
where $\beta_i$ are the projection coefficients.
The total intensity at the output is then
\begin{equation}
    I_\mathrm{R} = E_{\mathrm{R}}^{\dagger} E_{\mathrm{R}} = \frac{\sum_i s_i^2 |\beta_i|^2}{\sum_i|\beta_i|^2} \approx \langle s^2 \rangle \; \text{(weighted arithmetic mean).}
\end{equation}
Note that here the mean is computed over $m$ values.
For the global-focusing state, the output intensity is given by the product of two weighted arithmetic means, giving
\begin{equation}
    I_{\mathrm{G}} = \frac{\sum_i s_i^4 |\sum_j u^*_{j,i}|^2}{\sum_i s_i^2|\sum_j u^*_{j,i}|^2} \approx \frac{\langle s^4 \rangle }{\langle s^2 \rangle}.
\end{equation}
The latter equality is only approximate, as the $s_i$ are not statistically independent from $|\sum_j u^*_{j,i}|$.
Nevertheless, it allows for a good approximation of the enhancement which is given by the ratio of the global-focus intensity to the intensity obtained with a random input: $\eta_{\mathrm{I}}^\mathrm{G} = \frac{I_\mathrm{G}}{I_\mathrm{R}} = \frac{\langle s^4 \rangle }{\langle s^2 \rangle^2}$.
For the SVD the output intensity of the input vector $i$ is more straightforward to compute and is $I_i = s_i^2$, resulting in an enhancement $\eta_{\mathrm{I}}^{i} = \frac{s_i^2}{\langle s^2\rangle}$.
Now let us compare $\eta_{\mathrm{I}}^i$ obtained for the $i^{th}$ SVD vector and $\eta_{\mathrm{I}}^{\mathrm{G}}$,
\begin{equation}
    \eta_{\mathrm{I}}^{\mathrm{R}} = 1 \le \eta_{\mathrm{I}}^{\mathrm{G}} = \frac{ \langle s^4 \rangle}{ \langle s^2 \rangle ^2} \le \eta_{\mathrm{I}}^1 = \frac{s_1^2}{ \langle s^2 \rangle} = \tilde{s}_{1}^2.
    \label{eq:enhancement_comparison}
\end{equation}
The first inequality comes from Jensen's theorem and the second from the mean inequality.

\medskip

In the general, there is no obvious link between the intensity enhancement derived above and the field enhancement, which does not have a simple analytical derivation.
However, in case of Rayleigh statistics of the field one can construct this link.
One can show that for Rayleigh statistics the ratio of $\ell_1$ norms of two vectors is equal to the ratio of the $\ell_2$ norms of these two vectors.
Hence, because the amplitude enhancement corresponds to the vectors $\ell_1$ norms ratio and the intensity enhancement to the square $\ell_2$ norms ratio, one obtains
\begin{equation}
\eta_{\mathrm{E}} \simeq \sqrt{\eta_{\mathrm{I}}}=\tilde{s}.
\end{equation}
Note that this property does not hold for the amplitude enhancement of the global-focus due to its Rician statistics (see~\Sec{SM_speckle_stat}).
Finally, assuming a Marchenko-Pastur distribution, one can expect from \cref{eq:enhancement_comparison} that the intensity enhancements of the first singular vector scale as $1/\gamma$.
For the amplitude enhancements this results in a scaling with $1/\sqrt{\gamma}$ which corresponds well with the experimental observations presented in~\cref{fig:sum-a}.

\section{Speckle statistics}
\label{SM_speckle_stat}

\begin{figure}[h!]
    \phantomsubfloat{\label{fig:SM_stat-a}}
    \phantomsubfloat{\label{fig:SM_stat-b}}
    \includegraphics[width=0.6\columnwidth]{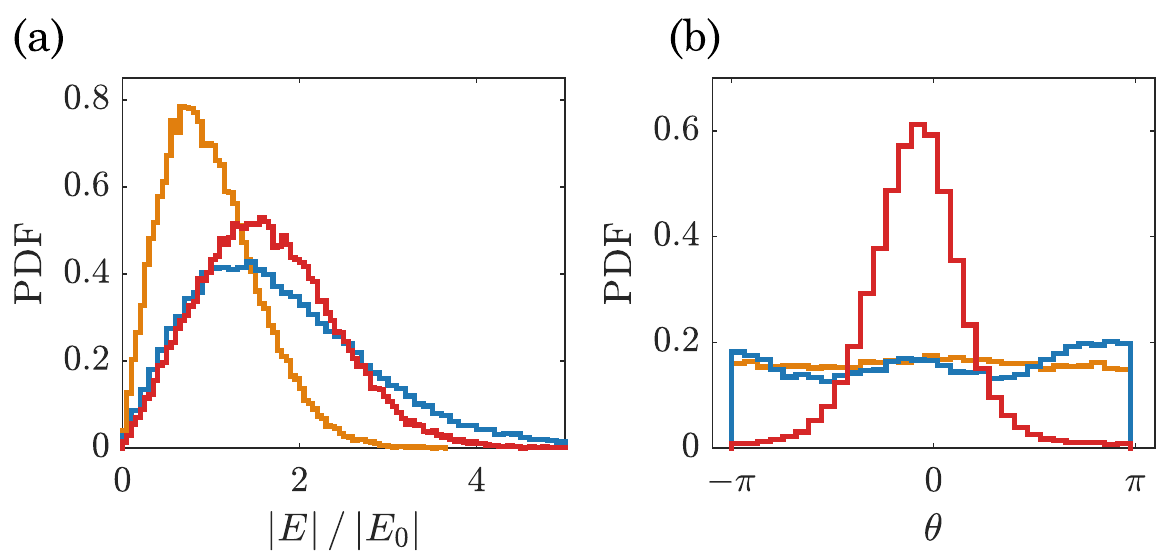}
    \caption{Speckle statistics.
    (a) Distribution of field amplitudes for three cases: field obtained from the first singular vector $v_1$ (blue), field of the global-focusing state (red) and a random reference input (yellow). All three distributions are normalized to the average field amplitude of the random reference $E_0$. (b) Corresponding phase distributions for the same data.}
    \label{fig:SM_stat}
\end{figure}

Fully developed speckles are governed by Rayleigh statistics: their amplitude is Rayleigh distributed while their phase distribution is flat.
While in the main text we primarily concentrated on the global modulation of the field amplitude at $\tau_0$, here we want to investigate the speckle distribution realized by the different input states.
\Cref{fig:SM_stat} shows that the reference field obtained for a random input as well as the different singular vectors reproduce the Rayleigh statistics ($v_1$ is shown as an example showing an enhanced average values compared to the reference).
However, the global-focusing pattern created by simultaneously focusing on each output pixel results in a Rician distribution of field amplitudes and a preferred phase~\cite{Goodman2007}.
This distribution corresponds to the sum of random phasors which have some common component while the Rayleigh distribution corresponds to the sum of fully random phasors.
The reason for the emergence of Rician statistics for the global-focus procedure is that it forces a common phase on all targeted output pixels. 

\end{document}